\documentclass[3p,authoryear]{elsarticle}
\journal{Planetary \& Space Science and now in press}

\usepackage{amssymb}
\usepackage{mathrsfs}
\usepackage{rotating}
\usepackage{graphicx}
\usepackage{epstopdf}
\usepackage{longtable,lscape}

\newcommand{\fn}[1]{\footnote{\scriptsize{#1}}} 

\newcommand{\Eqn}[1]{Eq{#1}.}  
\newcommand{\Fig}[1]{Fig{#1}.}  
\newcommand{\Cassit}{\textit{Cassini}}  
\newcommand{\ud}{\mathrm{d}}

\newcommand\aj{Astron.~J.~}%
\newcommand\apjl{Astrophys.~J.~Lett.~}%
\newcommand\aap{Astron.~Astrophys.~}%
\newcommand\grl{Geophys.~Res.~Lett.~}%
\newcommand\icarus{Icarus~}%
\newcommand\mnras{Mon.~Not.~Roy.~Astron.~Soc.~}%
\newcommand\nat{Nature~}%

\begin{document} 

\title{A modified ``Type~I migration'' model for\\propeller moons in Saturn's rings}

\author{Matthew~S.~Tiscareno}

\address{Center for Radiophysics and Space Research, Cornell University, Ithaca, NY 14853, USA.}

\begin{abstract}
We propose a mechanism for the observed non-keplerian motion \citep{Giantprops10} of ``propeller'' moons embedded in Saturn's rings.  Our mechanism, in which radial variations in surface density -- external to, and unaffected by, the embedded moon -- result in an equilibrium semimajor axis for the moon due to ``Type~I'' angular momentum exchange \citep{Crida10}, provides a good fit to the observations.  Future observations should distinguish between our model and others recently proposed. 
\end{abstract}

\maketitle

\section{Introduction}
A massive object embedded in a planetary ring creates a propeller-shaped disturbance in the local disk continuum \citep{SS00,SSD02,Seiss05}.  Swarms of such ``propellers,'' each with central moonlet radii~$r\sim100$~m, occur in the 3,000-km-wide ``Propeller Belts'' region in the middle part of Saturn's A~ring \citep{Propellers06,Propellers08,Sremcevic07}, though photometric ambiguities make it difficult to precisely ascertain the central moonlets' masses \citep{Anparsgw10}.  

\begin{figure*}[!t]
\begin{center}
\includegraphics[width=15cm,keepaspectratio=true]{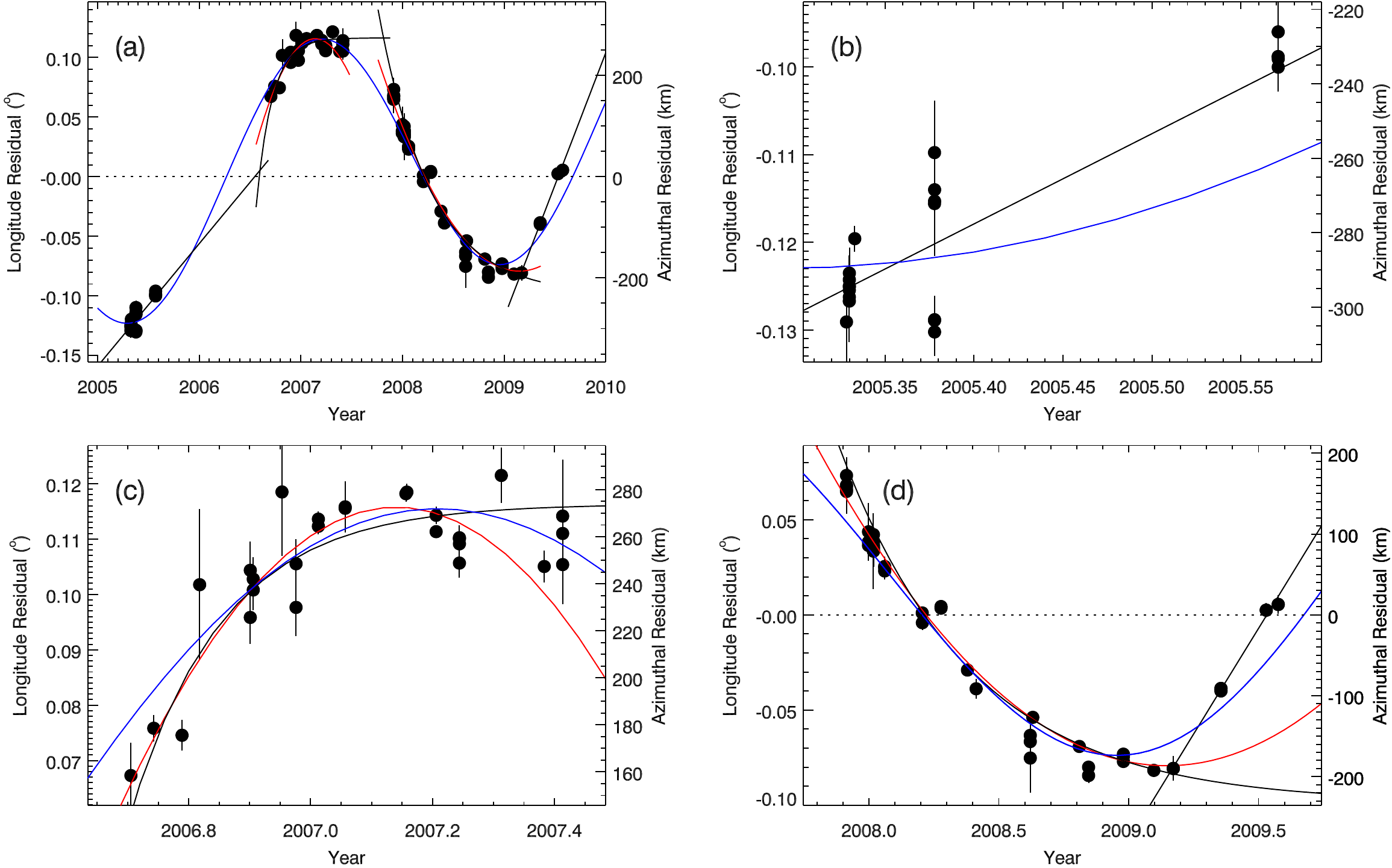}
\caption{Observed longitude of the propeller ``Bl\'eriot'' over 4~years, with linear trend $\dot{\lambda}_0=616.7819329^\circ$~day$^{-1}$ subtracted off.  Data reproduced from T10, q.v. for more detail.  Panel (a) contains all the data, while panels (b), (c), and (d) contain subsets of the data shown in greater detail.  The blue line indicates a linear-plus-sinusoidal fit to all the data, while the red lines indicate piecewise quadratic fits corresponding to a constant drift in semimajor axis (T10).  The black lines indicate exponential fits as described in Section~\ref{Analytical}.  The data from 2005 and 2009 are too sparse to distinguish among fitting functions, so are fit to a simple linear trend. 
\label{bleriot_orbit}}
\end{center}
\end{figure*}

A second propeller-rich region has been identified in the outermost regions of the A~ring \citep[hereafter T10]{Giantprops10}.  Propellers in this region are both larger and rarer than in the Propeller Belts, and several have been tracked for periods of several years.  The propeller nicknamed ``Bl\'eriot''\fn{Propellers that have been observed repeatedly are nicknamed to facilitate identification.} is the largest ($r\sim1$~km) and best-tracked ($>100$ detections over 5~years).  Its orbit is predominantly keplerian, with longitude residuals less than $\pm0.15^\circ$ (200~km) over a period of nearly 5~yr.  However, the residuals are much larger than measurement error and are clearly systematic (T10; data reproduced in \Fig{}~\ref{bleriot_orbit}).  If the observed non-keplerian motion is interpreted in terms of changes in the instantaneous semimajor axis $a$, then Bl\'eriot migrated outward from mid-2006 to mid-2007 at a rate of $\dot{a}=+0.11$~km/yr, and inward from late-2007 to early-2009 at a rate of $\dot{a}=-0.04$~km/yr (T10).  

A linear trend in semimajor axis, that is to say constant $\dot{a}$, corresponds to a quadratic trend in the longitude residual, $\lambda(t)\propto{}t^2$.  The observed longitude residuals for Bl\'eriot can be interpreted in terms of a series of piecewise quadratics (red lines in \Fig{}~\ref{bleriot_orbit}) that would imply episodic migration, perhaps caused and punctuated by periodic encounters or collisions \citep[T10]{LS09,Kirsh09}.  While plausible, this mechanism needs further development, and this paper will not discuss it further.  Other mechanisms for Bl\'eriot's non-keplerian motion have been proposed, and are reviewed below:  

The observed longitude residuals for Bl\'eriot can be interpreted in terms of a sinusoidal oscillation with period 3.7~yr (blue line in \Fig{}~\ref{bleriot_orbit}).  Although the most obvious physical mechanism to produce a sinusoidal longitude residual is resonant interaction with a larger moon exterior to the rings, a mechanism which governs many phenomena in the rings \citep[e.g.,][]{Ringschapter12}, no resonance is known near Bl\'eriot's position that could plausibly cause longitude residuals with the observed large amplitudes.  In the ``frog'' mechanism suggested by \citet[hereafter PC10]{PC10} and further developed by \citet{PC12}, the propeller moonlet interacts primarily with the mass at either end of the propeller gap.  Modeling those ends as co-orbiting masses, PC10 found that the propeller moonlet plausibly librates with the observed amplitude and period.  However, one wonders whether the moon-formed gap responds sluggishly enough to allow the moon to librate within it.  Two predictions of this may soon be tested: that Bl\'eriot's 3,000-km-long gap structure (which is only seen clearly in a few images) is stationary with respect to the librations, and that the longitude residual will continue to follow a sinusoidal profile into the future. 

\begin{figure*}[!t]
\begin{center}
\includegraphics[width=8cm,keepaspectratio=true]{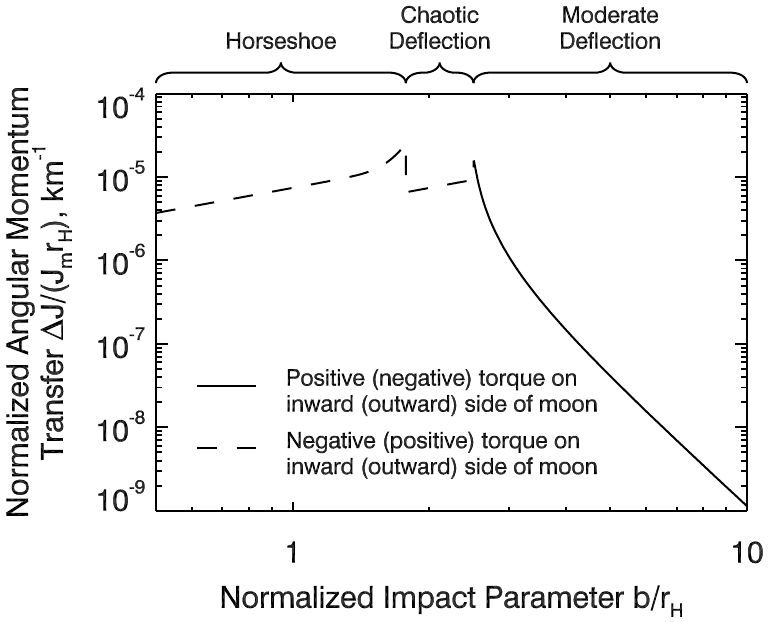}
\caption{Angular momentum exchange $\Delta{}J$ between a ring streamline and an embedded moon as a function of the normalized impact parameter $\hat{b}\equiv{}b/r_H$.  $\Delta{}J$ is normalized by the moon's specific angular momentum $J_m\equiv{}a_0^2n_0$ and by its Hill radius $r_H$.  Data from \citet{Crida10}. 
\label{cridacalc}}
\end{center}
\end{figure*}

Other hypotheses for Bl\'eriot's non-keplerian motion, including the one we present below, rely on the concept of ``Type~I migration.''  As classically formulated for protoplanetary disks \citep{Ward86,Ward97,PapaloizouPPV07}, the angular momentum exchange at inner Lindblad resonances between a disk and an embedded mass fails to exactly cancel with that at outer Lindblad resonances, resulting in a differential torque that leads to inward migration of the embedded mass.  However, classical Type~I migration depends crucially on the gas component of the disk, which causes Lindblad resonance locations to shift asymmetrically.  For the case of planetary rings, which are strictly particulate, \citet[hereafter C10]{Crida10} re-derived the equations for Type~I migration from first principles, using analytical arguments and numerical simulations to trace the angular momentum exchange between streamlines of continuum ring particles and the embedded moon, and deriving a profile of torque per unit disk surface density as a function of the impact parameter $b$ (i.e., the difference in semimajor axis) of ring particle streamlines with respect to the moon (\Fig{}~\ref{cridacalc}).  The primary contribution does not depend on resonances, which are more symmetrically placed in a non-gaseous particulate disk, but on intrinsic asymmetries in the single-impulse transfer of angular momentum. 

C10 find an intrinsic asymmetric torque, akin to classic Type~I migration, in the case of a particulate disk with uniform surface density; it is always inward (while Bl\'eriot is seen to move both inward and outward) and is one to two orders of magnitude too weak to explain the magnitude of Bl\'eriot's observed non-keplerian motion (C10).  However, it is known that Saturn's rings do not have uniform surface density.  \citet[hereafter RP10]{RP10} considered stochastic temporal variations of the surface density due to self-gravity wakes in the region of maximum angular momentum transfer, finding that they induce a random walk in the propeller moon's semimajor axis.  Under certain realistic conditions, the magnitude of the longitude residual approaches that observed for Bl\'eriot, and since a random walk has no preferred frequency, any frequency might be apparent over a limited period of time.  The model of RP10 predicts that the quasi-sinusoidal behavior observed by T10 will not continue, but will instead be seen as part of an overall random progression. 

\section{Analytical Model \label{Analytical}}
In Saturn's rings, the surface density changes not only with time (as in the model of RP10), but also with position (ring radius, or the distance from Saturn's center within the ring plane).  Thus, we propose here that Type~I torques can act to keep the embedded moon's semimajor axis at an equilibrium position, given a certain radial profile of the surface density.  Indeed, radial surface density structure in the ring means that any propeller must constantly migrate due to imbalanced torques, unless it finds a location where the torques balance.  Thus, we should expect to find them at such equilibrium points.

The ``sweet spots''\fn{In baseball, the ``sweet spot'' is the location on the bat at which a struck ball is most efficiently propelled.  So, here, we refer to the impact parameter at which angular momentum is most efficiently transferred between the moon and the disk.} where angular momentum exchange $|\Delta{}J(b)|$ is maximized in the C10 calculations of Type~I torque are at $\sim2$~Hill radii\fn{An object's Hill radius $r_H=a_0(m/3M)^{1/3}$, using variables defined later in this section, is the approximate extent of its gravitational dominance.} on either side of the embedded moon's semimajor axis (\Fig{}~\ref{cridacalc}).  These local maxima occur because $\Delta{}J(b)$ increases with impact parameter $b$ in the horseshoe region ($b\lesssim1.8r_H$), while it decreases with $b$ in the farther region ($b\gtrsim2.5r_H$) where particles are only moderately deflected by the encounter \citep[see, e.g., \Fig{}~3.30 of][]{MD99}.  Between those two regions lies a chaotic zone where $\Delta{}J(b)$ is depressed by the fact that particles can be deflected to either side of the moon, thus leading to two peaks at $b\sim1.8r_H$ and $b\sim2.5r_H$ (C10).  The sign of the torque is positive (negative) on the inward (outward) side of the moon in the moderate deflection region, and vice-versa for the horseshoe and chaotic regions (\Fig{}~\ref{cridacalc}).  

Although C10's results are plotted for only a single value of the embedded moon mass, scaling relations can be determined.  Such scaling is needed because the moon mass used in their models is 400~times larger than Bl\'eriot's likely mass (T10) 
-- in fact, it is 20~times larger than the moon Daphnis' mass \citep{PorcoSci07} 
and thus represents a moon that would not create a propeller structure but rather a fully cleared gap.  C10 state that their total cumulative torque scales as $m^{4/3}\propto{}r_H^4$, where $m$ is the mass of the embedded moon.  However,\fn{The insight of A.~Crida (personal communication, 2011) was invaluable in formulating the analysis given in the remainder of this paragraph.} one factor of $m^{1/3}\propto{}r_H$ does not enter into our calculations since it comes from integrating the difference between the torques on the inner and outer sides of the moon under C10's assumption of constant surface density, as their $\delta{}J(\hat{b})\equiv\Delta{}J(\hat{b})-\Delta{}J(-\hat{b})\propto\Delta{}J(\hat{b})\cdot\hat{b}r_H/a_0$ (\Eqn{}~34 of C10), using the scaled impact parameter $\hat{b}\equiv{}b/r_H$.  For our purposes, we note that $\Delta{}J(\hat{b})$ itself is proportional\fn{The asymptotic relations shown in \Fig{}~3 of C10 make this clear.  These expressions can be written as $\Delta{}J(\hat{b})\simeq(\hat{b}r_H/a_0)J_m$ for the horseshoe region and $\Delta{}J(\hat{b})\simeq(\hat{b}r_H/2a_0)J_m$ for the chaotic deflection region.  For the moderate deflection region, $\Delta{}J(\hat{b})$ as expressed in \Eqn{}~32 of C10 is proportional to $m^2/b^5\propto{}r_H/\hat{b}^5$.  The inference that $\Delta{}J(\hat{b})\propto{}r_H$ is confirmed in direct calculations by A.~Crida (personal communication, 2011).}
 to $J_mr_H$, where the moon's specific angular momentum is $J_m\equiv{}a_0^2n_0$ for semimajor axis $a_0$ and mean motion $n_0$ (for this reason, in \Fig{}~\ref{cridacalc}, we normalize $\Delta{}J(\hat{b})$ by $r_H$ as well as by $J_m$, obtaining a curve that applies to embedded moons of any mass), and that a further factor of $r_H^2$ arises when calculating the integrated torque (rewriting \Eqn{}~36 of C10):
\begin{equation}
\label{CridaEqn36}
T=3n_0r_H^2\int_{-\infty}^\infty\sigma(\hat{b}')\Delta{}J(\hat{b}')\hat{b}'\ud\hat{b}'. 
\end{equation}
\noindent Thus, for our case in which $\sigma(\hat{b})$ must be convolved with the angular momentum exchange $\Delta{}J(\hat{b})$, rather than being held as a constant so that $\delta{}J(\hat{b})$ becomes important, the total torque $T\propto{}r_H^3\propto{}m$.  Since migration rate $\dot{a}\propto{}T/m$, we conclude that a given variation in surface density will induce the same $\dot{a}$ in moons of different masses, though the radial scale over which the surface density variation occurs must scale with $r_H$ -- so that $\sigma(\hat{b})$ remains constant -- to preserve the effect. 

Consider a ring whose surface density varies radially but is azimuthally symmetric, and which has a local minimum in surface density at a given radius, as in \Fig{}~\ref{MovieInwardDeltafunc}.  The radial density structure $\sigma(b)$ is external to, and unaffected by, the embedded moon; in fact, such radial density structure is common in Saturn's rings (see Section~\ref{Numerical} for more discussion).  The moon's evolution is primarily driven by variations in $\sigma(b)$ at the ``sweet spots'' on either side of the moon.  The total torque $T$ on the moon is found by integrating \Eqn{}~\ref{CridaEqn36}, convolving $\Delta{}J(b)$ with $\sigma(b)$; a stable equilibrium can be created simply by setting $\ud\sigma/\ud{}b$ negative on the inward side of the moon and positive on the outward side.  If the embedded moon has its semimajor axis suddenly altered from this equilibrium position by an encounter \citep{LS09}, for example with a large fluctuation of self-gravity wakes or with a large ring particle, the torque from the surrounding material will then be out of balance and will push the moon's semimajor axis back toward the equilibrium position (\Fig{}~\ref{MovieInwardDeltafunc}). 

In the following paragraphs we analytically consider the simplest case, in which $\ud\sigma/\ud{}b$ has a constant negative value at the inward sweet spot, and the same constant value (except positive) at the outward sweet spot, and the equilibrium semimajor axis $a_0$ lies exactly between the two regions.  If we simplify the angular momentum exchange profile as two delta functions, one on each side of the moon,
\begin{equation}
\label{DeltaFunc}
\frac{\ud{}T}{\sigma}=A\delta(b\pm{}b_0),
\end{equation}
\noindent where $A$ and $b_0$ are free parameters (we use $A=4.7\times10^{18}$~cm$^4$~s$^{-2}$, a value obtained by integrating \Eqn{}~\ref{CridaEqn36} with $\Delta{}J(\hat{b})$ given by \Fig{}~\ref{cridacalc} and $\sigma(\hat{b})$ set equal to unity; and $b_0=2.78$~km in order to mimic the balance point of the full exchange function), 
then the convolved torque is given by
\begin{equation}
\label{ToyModelTorque}
T=-2A\frac{\ud\sigma}{\ud{}r}(a-a_0),
\end{equation}
\noindent where $a_0$ is the equilibrium semimajor axis and $a$ is the moon's actual semimajor axis.  The angular momentum is given by $L=m(GMa)^{1/2}$, assuming that any non-zero eccentricity will be quickly damped away, where $G$ is Newton's constant and $M$ and $m$ are the masses of Saturn and the moonlet.  We take the latter as $m\simeq2\times10^{15}\mathrm{~g}\simeq4\times10^{-15}M$ for an icy moon with radius $\sim1$~km (T10) and internal density close to the Roche critical density $\rho_c\approx0.5$~g~cm$^{-3}$ \citep{PorcoSci07}.  
The torque is then $T=\dot{L}=mn_0a_0\dot{a}/2$. 
\noindent Setting this equal to \Eqn{}~\ref{ToyModelTorque}, we obtain a simple differential equation for the semimajor axis offset $a-a_0$:
\begin{equation}
\frac{\ud(a-a_0)}{\ud{}t}+\omega_0(a-a_0)=0,
\end{equation}
\noindent where
\begin{equation}
\label{Omega0Eqn}
\omega_0\equiv\frac{4A}{mn_0a_0}\frac{\ud\sigma}{\ud{}r}.  
\end{equation}
\noindent The solution, with initial displacement $a=a_d$ at $t=0$, is
\begin{equation}
\label{ToyModelSoln}
(a-a_0)=(a_d-a_0)e^{-\omega_0t}.  
\end{equation}
\noindent Using the standard equation for keplerian shear and substituting \Eqn{}~\ref{ToyModelSoln}, the longitude residual $\lambda-\lambda_0$, where $\lambda_0\equiv{}n_0t$, will accumulate as
\begin{equation}
\frac{\ud (\lambda-\lambda_0)}{\ud t} = - \frac{3}{2} \frac{n_0}{a_0} (a_d-a_0) e^{-\omega_0 t} .  
\end{equation}
\noindent Integrating, we obtain
\begin{equation}
\label{LongResidEqn}
\lambda-\lambda_0=\frac{3}{2}\frac{n_0}{a_0\omega_0}(a_d-a_0)\left(e^{-\omega_0t}-1\right).
\end{equation}
\noindent Note that, as $t\to\infty$, the longitude residual asymptotically approaches a constant negative (positive) value for a positive (negative) initial offset in semimajor axis. 

\begin{table*}[!t]
\caption{Fit parameters to $\lambda-n_0t=p_1+(p_0-p_1)e^{-2\pi(t-p_3)/p_2}$ for data in \Fig{}~\ref{bleriot_orbit}. 
\label{propeller_efits}}
\begin{scriptsize}
\begin{tabular}{l c c c c}
\hline
\hline
Year & $p_0$ & $p_1$ & $p_2$ & $p_3$ \\
\hline
mid-2006 to early-2007 & 0.00 & 0.12 & 0.98 & 2006.59 \\
late-2007 to early-2009 & 0.066 & -0.098 & 3.21 & 2007.94 \\
\hline
\end{tabular}
\end{scriptsize}
\end{table*}

This simplified model result can be compared to data.  In \Fig{}~\ref{bleriot_orbit}, the observed longitude residuals in two well-sampled periods are separately fit to functions with the form of \Eqn{}~\ref{LongResidEqn} (see black lines in \Fig{}~\ref{bleriot_orbit}, and Table~\ref{propeller_efits}).  
In the interval from mid-2006 to early-2007 (\Fig{}~\ref{bleriot_orbit}c), we obtained $(2\pi/\omega_0)=1.0$~yr with $\lambda-\lambda_0=270\mathrm{~km}=0.0020$~radians.  
Thus, taking $n_0=616.782^\circ$~day$^{-1}$ at $a_0=134912$~km (T10), we calculate from \Eqn{}~\ref{Omega0Eqn} that $\ud\sigma/\ud{}b=3.8$~g~cm$^{-2}$~km$^{-1}$
and from \Eqn{}~\ref{LongResidEqn} that $a_d-a_0=-300$~m. 
Similarly, in the interval from late-2007 to early-2009 (\Fig{}~\ref{bleriot_orbit}d), we obtained $(2\pi/\omega_0)=3.2$~yr with $\lambda-\lambda_0=-390\mathrm{~km}=-0.0029$~radians, 
yielding $\ud\sigma/\ud{}b=1.2$~g~cm$^{-2}$~km$^{-1}$
and $a_d-a_0=130$~m. 

The initiation of our mechanism is not simply explained.  Our model requires three events (where the black lines cross in \Fig{}~\ref{bleriot_orbit}) over our five-year data set at which there is a rapid change in $a$ on the order of a few $\times 100$~m.  Direct delivery of sufficient momentum could occur if an object with approximately one-thirtieth the mass of Bl\'eriot 
(i.e., one-third its radius) 
strikes at Bl\'eriot's escape velocity of $\sim50$~cm~s$^{-1}$, 
The size distribution reported by T10 predicts that approximately one such object should occur per annulus of radial width 35~km, 
while the radial cross-section for interacting with Bl\'eriot is at least the diameter of its central moonlet $\sim2$~km and more likely the total radial width of the cleared region $\sim15$~km -- this is the sum of the radial offset between the two lobes $\Delta{}r\sim5$~km (T10) and the radial full-width at half-maximum of a lobe, which we estimate from the T10 dataset to be $\sim10$~km. 
Another possibility is interaction with an unusually dense ($\sim$once-per-year) self-gravity wake (SGW; see \cite{Cuzzi10}) of similar mass.  Little is known about the detailed dynamics of SGWs, especially infrequent outliers.  Any resulting eccentricity need not be quickly damped; there is no empirical evidence to discount a small eccentricity $\sim{}10^{-6}$ for Bl\'eriot, which would lead to azimuthal residuals $\sim{}0.1$~km.

We also note that the method of \cite{RP10}, which has kicks from SGWs at its core while we mention such kicks above only as a possible initiation for our mechanism that focuses on disk-moon torques, similarly suffers from problems with the magnitude of the available perturbations.  

However, we believe this model remains worthy of consideration because the data (\Fig{}~\ref{bleriot_orbit}) resemble a series of curves somewhat better than they resemble a smooth sinusoid.  The residuals for this new functional fit are significantly better than for the sinusoidal fit (though with 10 degrees of freedom, compared with 5 for the sinusoidal fit), and are comparable to those for the piecewise quadratic fit (T10), with particular improvement for the spring 2007 data (\Fig{}~\ref{bleriot_orbit}c).  

The present model requires only three ``kick'' events to explain the 5-year data set, since the 2006/07 and the 2007/09 lines intersect each other.  However, the differing values of $\omega_0$ for the two segments may indicate that $\ud\sigma/\ud{}b$ does not have the same value at the inward and outward ``sweet spots'', leading to differing response times to an offset.  With this in mind, we construct a simple numerical version of our model in the next section, using the entire angular momentum exchange profile rather than reducing it to a pair of delta functions as we did for the above analytical treatment. 

\section{Numerical Model \label{Numerical}}

\begin{figure*}[!t]
\begin{center}
\includegraphics[width=15cm,keepaspectratio=true]{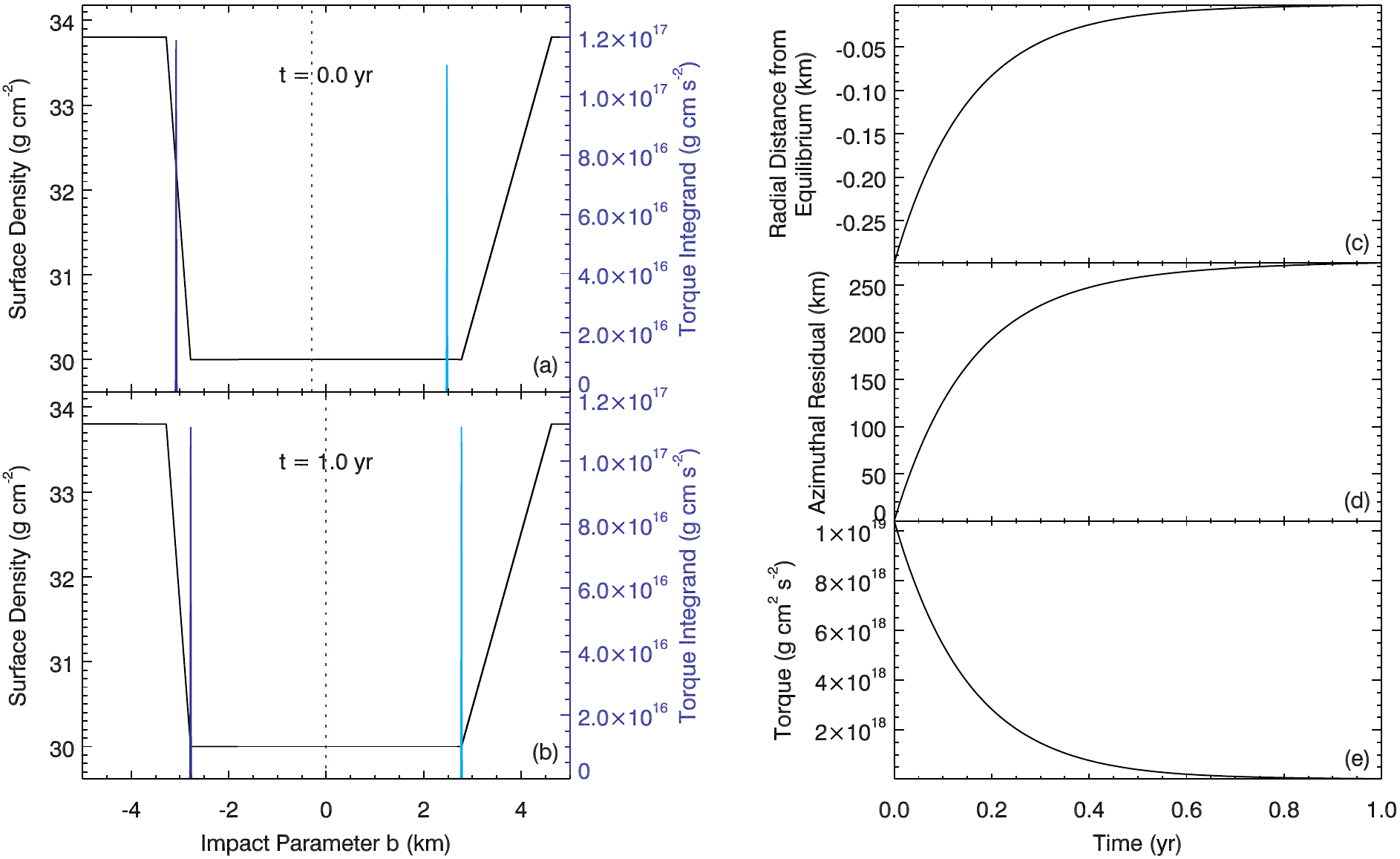}
\caption{In this toy model, the embedded moon's semimajor axis (dotted blue line) begins with an initial inward displacement from equilibrium at $t=0$ (a) and trends back to its equilibrium value (b).  The blue and cyan curves are the angular momentum exchange profile (positive and negative contributions, respectively), here approximated as a pair of narrow gaussians (\Eqn{}~\ref{DeltaFunc}).  The black curve is an input radial surface density ($\sigma$) profile, with $\ud\sigma/\ud{}b=-4.8$~g~cm$^{-2}$~km$^{-1}$ on the inward side of the equilibrium semimajor axis and $+1.44$~g~cm$^{-2}$~km$^{-1}$ on the outward side.  Both slopes occur only in the region of maximum angular momentum exchange.  The right-hand panels show (c) the semimajor axis offset $\Delta{}a$, (d) the azimuthal residual $\Delta\lambda$, and (e) the torque $T$ as functions of time. 
\label{MovieInwardDeltafunc}}
\end{center}
\end{figure*}

We verified the above analytical results by numerically integrating the C10 equations.  

Firstly, we represented the angular momentum exchange function as a pair of narrow gaussians after the manner of \Eqn{}~\ref{DeltaFunc}.  In two separate runs, with a radial surface density profile consisting of a ``valley'' with slopes of $\ud\sigma/\ud{}b=3.8$~(1.2)~g~cm$^{-2}$~km$^{-1}$ in the vicinity of $b_0$, and with an initial semimajor axis $-300\mathrm{~}(+130)$~m from the e, we indeed found that the semimiajor axis took about 1.0~(3.2)~yr to return to equilibrium and that the total azimuthal residual came to about $+270\mathrm{~(-390)}$~km.  

\begin{figure*}[!t]
\begin{center}
\includegraphics[width=12.2cm,keepaspectratio=true]{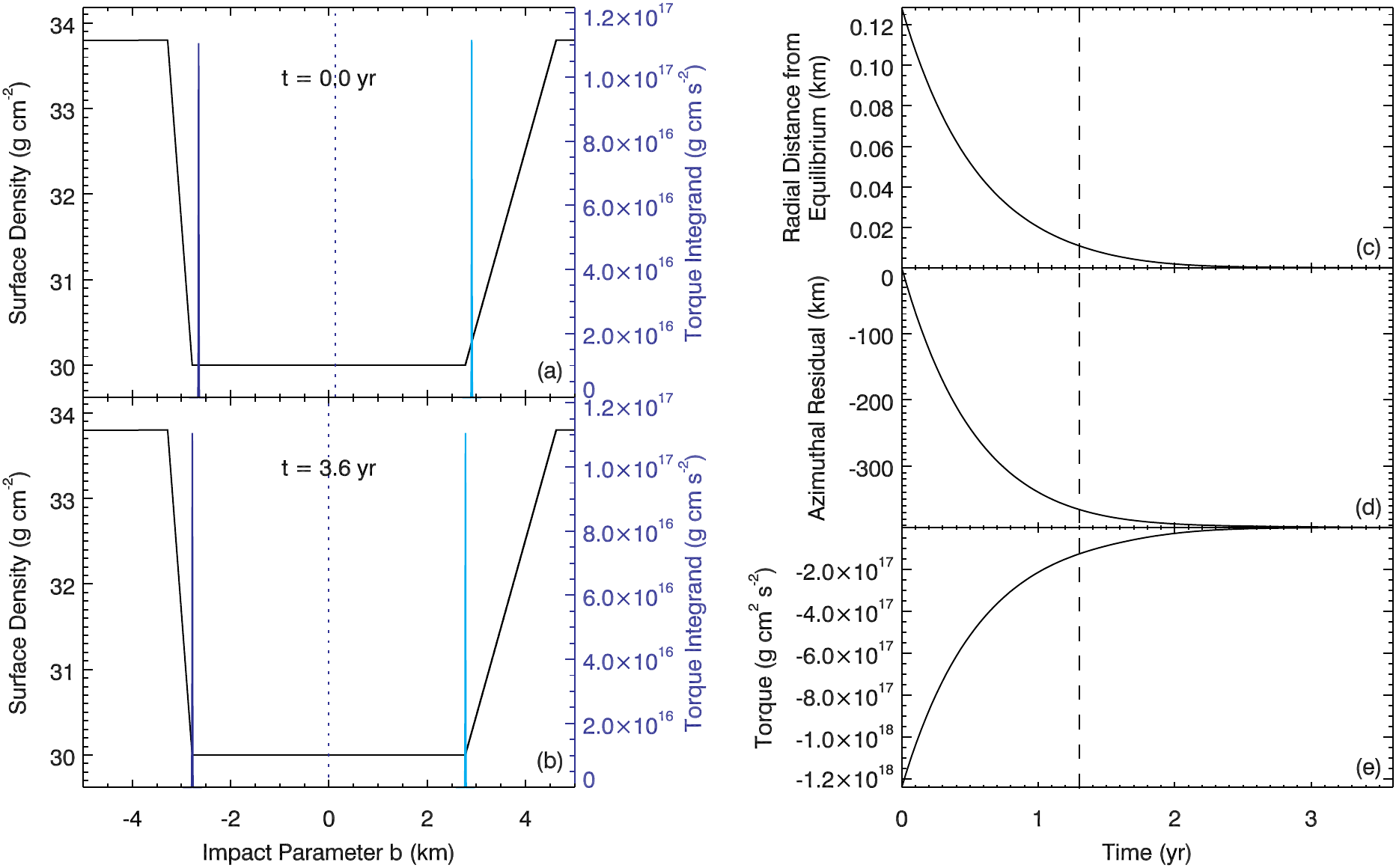}
\caption{This figure is identical to \Fig{}~\ref{MovieInwardDeltafunc}, but for an outward initial displacement.  The vertical dashed line in panels (c), (d), and (e) indicates the approximate length of the observation interval (\Fig{}~\ref{bleriot_orbit}d). 
\label{MovieOutwardDeltafunc}}
\end{center}
\end{figure*}

Secondly, to produce a model that does not invoke temporal changes in the surface density profile, we used a profile with a steeper slope on the inward side and a gentler slope on the outward side (\Fig{s}~\ref{MovieInwardDeltafunc} and~\ref{MovieOutwardDeltafunc}).  In this way, only the inward (outward) slope is encountered by the exchange functions when the semimajor axis is offset inward (outward), and different response times for the the two cases are thus possible.  We again obtained results similar to the observations when the ``valley'' floor was 3.8~g~cm$^{-2}$ below the background.

\begin{figure*}[!t]
\begin{center}
\includegraphics[width=15cm,keepaspectratio=true]{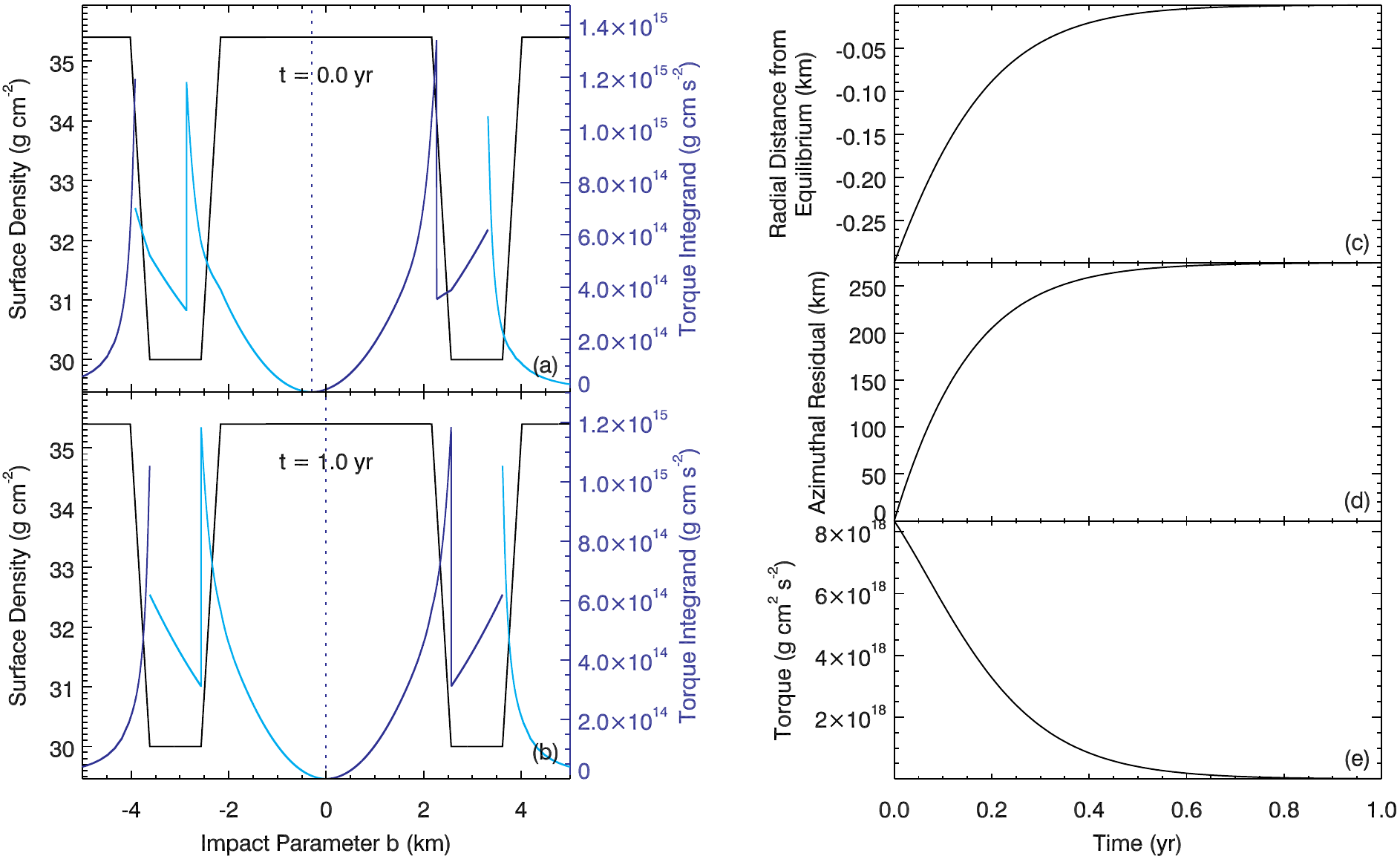}
\caption{The same as \Fig{}~\ref{MovieInwardDeltafunc}, but with the full angular momentum exchange profile (the profile shown in \Fig{}~\ref{cridacalc} multipled by the surface density profile), and with a double-troughed surface-density profile with $\ud\sigma/\ud{}b=\pm$~13.5~g~cm$^{-2}$~km$^{-1}$ in regions of maximum angular momentum exchange.  
\label{MovieInward}}
\end{center}
\end{figure*}

\begin{figure*}[!b]
\begin{center}
\includegraphics[width=12.2cm,keepaspectratio=true]{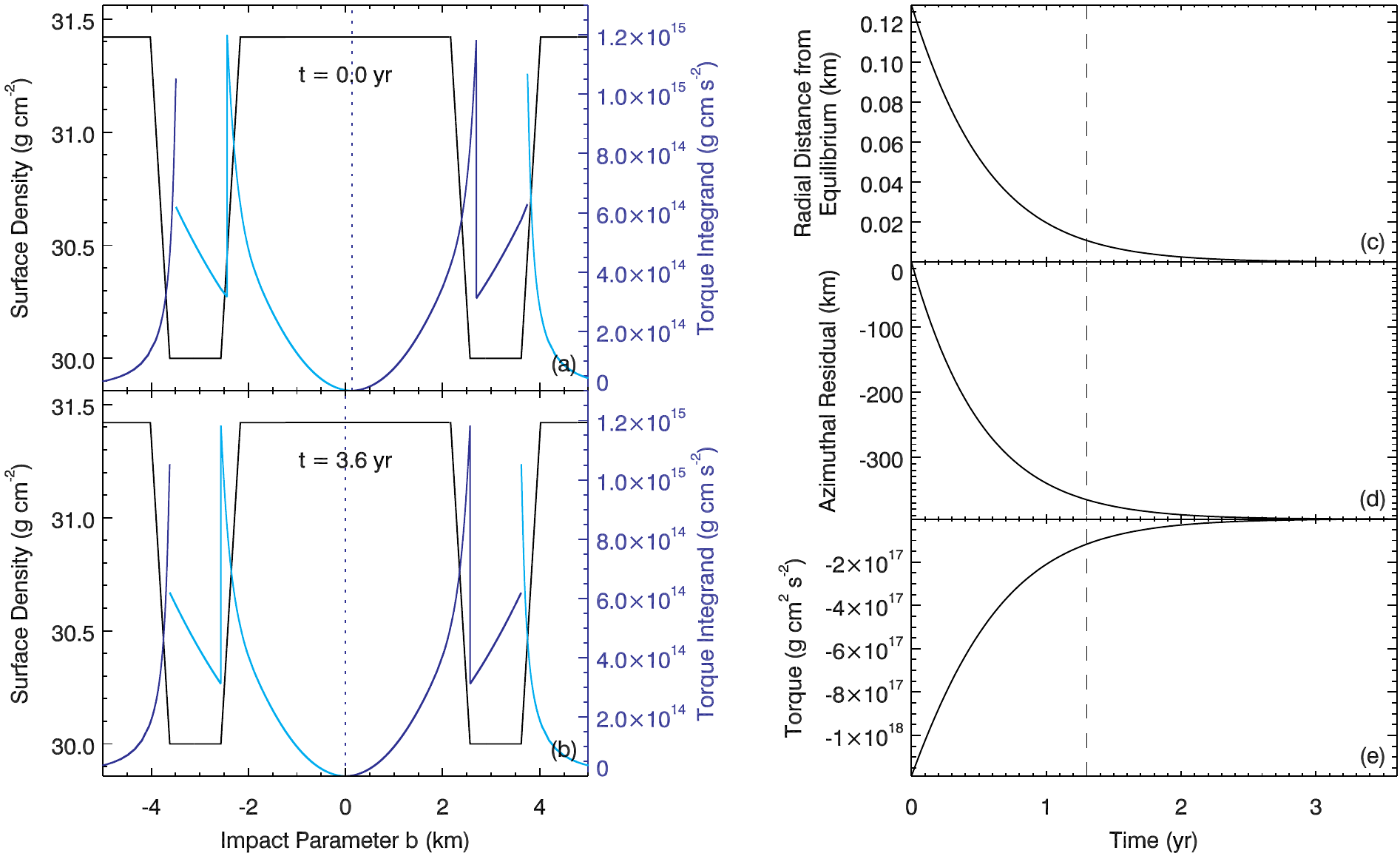}
\caption{This figure is identical to \Fig{}~\ref{MovieInward}, but for an outward initial displacement and with $\ud\sigma/\ud{}b=\pm$~3.55~g~cm$^{-2}$~km$^{-1}$.  The vertical dashed line in panels (c), (d), and (e) indicates the approximate length of the observation interval (\Fig{}~\ref{bleriot_orbit}d). 
\label{MovieOutward}}
\end{center}
\end{figure*}

Finally, we extend our treatment to the case of the actual angular momentum exchange function (\Fig{}~\ref{cridacalc}).  There are many ways to configure the surface density so that the longitude residual matches the data, including a square well and the modified square well seen in \Fig{s}~\ref{MovieInwardDeltafunc} and~\ref{MovieOutwardDeltafunc}, but we found the most efficient model (requiring the smallest $\Delta\sigma$) to be that shown in \Fig{s}~\ref{MovieInward} and~\ref{MovieOutward}.  Because the angular momentum exchange is spread over a larger annulus, it is harder to have steeper slopes encountered only for an inward initial offset and gentler slopes only for an outward initial offset, as in the previous paragraph.  In the example shown, the ``valleys'' are 5.4~(1.42)~g~cm$^{-2}$ lower than the background surface density for initial semimajor axis offset in the inward (outward) direction, yielding slopes of $\ud\sigma/\ud{}b=13.5$~(3.55)~g~cm$^{-2}$~km$^{-2}$.  

\section{Discussion and Conclusions \label{Discussion}}

\begin{figure}[!t]
\begin{center}
\includegraphics[width=13.25cm,keepaspectratio=true]{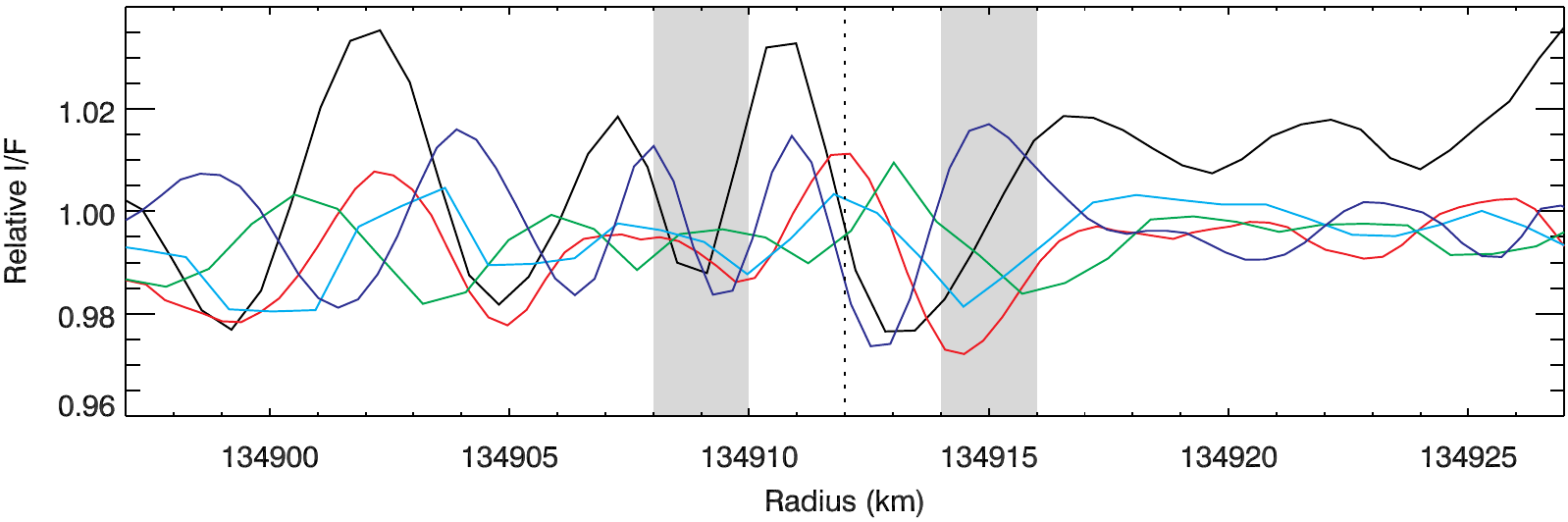}
\caption{Normalized brightness ($I/F$) profiles of the A~ring in the vicinity of Bl\'eriot's semimajor axis (vertical dotted line) from the following images:  N1540681631, taken on 2006~October~27 (green); N1541717040, taken on 2006~November~8 (cyan); N1560309735, taken on 2007~June~12 (blue); N1595337202, taken on 2008~July~21 (red); and N1654254507, taken on 2010~June~3 (black).  Regions of maximum angular momentum exchange are shaded in gray. 
\label{typeiprops_data}}
\end{center}
\end{figure}

We have shown that a modest permanent radial variation in the surface density, combined with plausible occasional ``kicks'' to the semimajor axis, can lead to non-keplerian motions of an embedded moonlet that are similar to those observed for Bl\'eriot.  It finally remains to inquire whether such variations may actually exist in Bl\'eriot's vicinity.  This region is characterized by variable structure (\Fig{}~\ref{typeiprops_data}), most likely due to spiral density waves.  Our model is viable if, when averaged over orbital timescales (which are short compared to the migration rate), the mean torques due to that structure yield anomalies on either side of Bl\'eriot's position.  Optical-depth variations in this part of the A~ring are $\pm0.2$ (M.~Hedman, personal communication, 2011).  Using a mass extinction coefficient for the outer-A~ring of 0.01--0.02~cm$^2$~g$^{-1}$ \citep{Colwell09}, this corresponds to surface density variations as high as $\pm10$~g~cm$^{-2}$.
The mechanism described herein requires surface density anomalies of only a few g~cm$^{-2}$.

Although we cannot prove whether the observed surface density variations are underlain by the much smaller average anomalies necessary to activate our mechanism, nor is such proof likely to be possible without greatly improved observations and/or difficult numerical modeling, our purpose here is only to argue that it is plausible.  The model presented in this paper is motivated by the observed orbital evolution of Bl\'eriot and other propellers, which seems to be characterized by periodic ``kicks,'' with slow variations in the semimajor axis (seen as curvature in the longitude residuals) occurring between those kicks.  It is in response to these observations that we have constructed a mechanism that is dynamically plausible and consistent with existing data. 

Our proposed mechanism joins two others recently outlined, and all three mechanisms make contrasting predictions regarding the future behavior of Bl\'eriot's azimuthal residual.  While PC10 predict that the quasi-sinusoidal trend discerned in the azimuthal residual to date will continue, and RP10 predict that variations in the azimuthal residual will be seen in the future to result from stochastic variation in the semimajor axis, we predict periodic (less than one per year, on average, from our interpretation of existing data) sudden ``kicks'' to the semimajor axis followed by asymptotic return to the equilibrium value.  Future data, beginning in 2013 when \Cassit{} will again spend significant time out of Saturn's equatorial plane with good viewing of the rings, should distinguish among these models. 
\\
\\
\textbf{Acknowledgements} I thank J.~Burns, M.~Hedman, P.~Nicholson, D.~Hamilton, M.~Evans, A.~Crida, and a reviewer for helpful comments, and I further thank A.~Crida for sharing his calculation results.  I acknowledge funding from NASA's Outer Planets Research program (NNX10AP94G).  This paper is dedicated to~CGT.


\begin{thebibliography}{22}
\expandafter\ifx\csname natexlab\endcsname\relax\def\natexlab#1{#1}\fi
\expandafter\ifx\csname url\endcsname\relax
  \def\url#1{\texttt{#1}}\fi
\expandafter\ifx\csname urlprefix\endcsname\relax\def\urlprefix{URL }\fi

\bibitem[{{Colwell} et~al.(2009){Colwell}, {Cooney}, {Esposito}, and {Srem{\v
  c}evi{\'c}}}]{Colwell09}
{Colwell}, J.~E., {Cooney}, J.~H., {Esposito}, L.~W., {Srem{\v c}evi{\'c}}, M.,
  2009. {Density waves in Cassini UVIS stellar occultations. 1. The Cassini
  Division}. \icarus  200, 574--580.

\bibitem[{{Crida} et~al.(2010){Crida}, {Papaloizou}, {Rein}, {Charnoz}, and
  {Salmon}}]{Crida10}
{Crida}, A., {Papaloizou}, J.~C.~B., {Rein}, H., {Charnoz}, S., {Salmon}, J.,
  2010. {Migration of a moonlet in a ring of solid particles: Theory and
  application to Saturn's propellers}. \aj  140, 944--953.

\bibitem[{{Cuzzi} et~al.(2010){Cuzzi}, {Burns}, {Charnoz}, {Clark}, {Colwell},
  {Dones}, {Esposito}, {Filacchione}, {French}, {Hedman}, {Kempf}, {Marouf},
  {Murray}, {Nicholson}, {Porco}, {Schmidt}, {Showalter}, {Spilker}, {Spitale},
  {Srama}, {Srem{\v c}evi{\'c}}, {Tiscareno}, and {Weiss}}]{Cuzzi10}
{Cuzzi}, J.~N., {Burns}, J.~A., {Charnoz}, S., {Clark}, R.~N., {Colwell},
  J.~E., {Dones}, L., {Esposito}, L.~W., {Filacchione}, G., {French}, R.~G.,
  {Hedman}, M.~M., {Kempf}, S., {Marouf}, E.~A., {Murray}, C.~D., {Nicholson},
  P.~D., {Porco}, C.~C., {Schmidt}, J., {Showalter}, M.~R., {Spilker}, L.~J.,
  {Spitale}, J.~N., {Srama}, R., {Srem{\v c}evi{\'c}}, M., {Tiscareno}, M.~S.,
  {Weiss}, J., 2010. {An evolving view of Saturn's dynamic rings}. Science
  327, 1470--1475.

\bibitem[{{Kirsh} et~al.(2009){Kirsh}, {Duncan}, {Brasser}, and
  {Levison}}]{Kirsh09}
{Kirsh}, D.~R., {Duncan}, M., {Brasser}, R., {Levison}, H.~F., 2009.
  {Simulations of planet migration driven by planetesimal scattering}. \icarus
  199, 197--209.

\bibitem[{{Lewis} and {Stewart}(2009)}]{LS09}
{Lewis}, M.~C., {Stewart}, G.~R., 2009. {Features around embedded moonlets in
  Saturn's rings: The role of self-gravity and particle size distributions}.
  \icarus  199, 387--412.

\bibitem[{{Murray} and {Dermott}(1999)}]{MD99}
{Murray}, C.~D., {Dermott}, S.~F., 1999. {Solar System Dynamics}. Cambridge
  Univ. Press, Cambridge.

\bibitem[{{Pan} and {Chiang}(2010)}]{PC10}
{Pan}, M., {Chiang}, E., 2010. {The propeller and the frog}. \apjl  722,
  L178--L182.

\bibitem[{{Pan} and {Chiang}(2012)}]{PC12}
{Pan}, M., {Chiang}, E., 2012. {Care and feeding of frogs}. \aj  143, 9.

\bibitem[{{Papaloizou} et~al.(2007){Papaloizou}, {Nelson}, {Kley}, {Masset},
  and {Artymowicz}}]{PapaloizouPPV07}
{Papaloizou}, J.~C.~B., {Nelson}, R.~P., {Kley}, W., {Masset}, F.~S.,
  {Artymowicz}, P., 2007. {Disk-planet interactions during planet formation}.
  In: {Reipurth}, B., {Jewitt}, D., {Keil}, K. (Eds.), Protostars and Planets
  V. Univ. Arizona Press, Tucson, pp. 655--668.

\bibitem[{{Porco} et~al.(2007){Porco}, {Thomas}, {Weiss}, and
  {Richardson}}]{PorcoSci07}
{Porco}, C.~C., {Thomas}, P.~C., {Weiss}, J.~W., {Richardson}, D.~C., 2007.
  {Saturn's small satellites: Clues to their origins}. Science  318,
  1602--1607.

\bibitem[{{Rein} and {Papaloizou}(2010)}]{RP10}
{Rein}, H., {Papaloizou}, J.~C.~B., 2010. {Stochastic orbital migration of
  small bodies in Saturn's rings}. \aap  524, A22.

\bibitem[{{Sei{\ss}} et~al.(2005){Sei{\ss}}, {Spahn}, {Srem{\v c}evi{\'c}}, and
  {Salo}}]{Seiss05}
{Sei{\ss}}, M., {Spahn}, F., {Srem{\v c}evi{\'c}}, M., {Salo}, H., 2005.
  {Structures induced by small moonlets in Saturn's rings: Implications for the
  Cassini mission}. \grl  32, L11205.

\bibitem[{{Spahn} and {Srem{\v c}evi{\'c}}(2000)}]{SS00}
{Spahn}, F., {Srem{\v c}evi{\'c}}, M., 2000. {Density patterns induced by small
  moonlets in Saturn's rings?} \aap  358, 368--372.

\bibitem[{{Srem{\v c}evi{\'c}} et~al.(2007){Srem{\v c}evi{\'c}}, {Schmidt},
  {Salo}, {Sei{\ss}}, {Spahn}, and {Albers}}]{Sremcevic07}
{Srem{\v c}evi{\'c}}, M., {Schmidt}, J., {Salo}, H., {Sei{\ss}}, M., {Spahn},
  F., {Albers}, N., 2007. {A belt of moonlets in Saturn's A ring}. \nat  449,
  1019--1021.

\bibitem[{{Srem{\v c}evi{\'c}} et~al.(2002){Srem{\v c}evi{\'c}}, {Spahn}, and
  {Duschl}}]{SSD02}
{Srem{\v c}evi{\'c}}, M., {Spahn}, F., {Duschl}, W.~J., 2002. {Density
  structures in perturbed thin cold discs}. \mnras  337, 1139--1152.

\bibitem[{{Tiscareno}(2012)}]{Ringschapter12}
{Tiscareno}, M.~S., 2012. {Planetary rings}. In: {French}, L., {Kalas}, P.
  (Eds.), Solar and Planetary Systems. Springer, Dordrecht, in press
  (arXiv:1112.3305).

\bibitem[{{Tiscareno} et~al.(2008){Tiscareno}, {Burns}, {Hedman}, and
  {Porco}}]{Propellers08}
{Tiscareno}, M.~S., {Burns}, J.~A., {Hedman}, M.~M., {Porco}, C.~C., 2008. {The
  population of propellers in Saturn's A ring}. \aj  135, 1083--1091.

\bibitem[{{Tiscareno} et~al.(2006){Tiscareno}, {Burns}, {Hedman}, {Porco},
  {Weiss}, {Dones}, {Richardson}, and {Murray}}]{Propellers06}
{Tiscareno}, M.~S., {Burns}, J.~A., {Hedman}, M.~M., {Porco}, C.~C., {Weiss},
  J.~W., {Dones}, L., {Richardson}, D.~C., {Murray}, C.~D., 2006.
  {100-metre-diameter moonlets in Saturn's A Ring from observations of
  ``propeller'' structures}. \nat  440, 648--650.

\bibitem[{{Tiscareno} et~al.(2010{\natexlab{a}}){Tiscareno}, {Burns}, {Srem{\v
  c}evi{\'c}}, {Beurle}, {Hedman}, {Cooper}, {Milano}, {Evans}, {Porco},
  {Spitale}, and {Weiss}}]{Giantprops10}
{Tiscareno}, M.~S., {Burns}, J.~A., {Srem{\v c}evi{\'c}}, M., {Beurle}, K.,
  {Hedman}, M.~M., {Cooper}, N.~J., {Milano}, A.~J., {Evans}, M.~W., {Porco},
  C.~C., {Spitale}, J.~N., {Weiss}, J.~W., 2010{\natexlab{a}}. {Physical
  characteristics and non-keplerian orbital motion of ``propeller'' moons
  embedded in Saturn's rings}. \apjl  718, L92--L96.

\bibitem[{{Tiscareno} et~al.(2010{\natexlab{b}}){Tiscareno}, {Perrine},
  {Richardson}, {Hedman}, {Weiss}, {Porco}, and {Burns}}]{Anparsgw10}
{Tiscareno}, M.~S., {Perrine}, R.~P., {Richardson}, D.~C., {Hedman}, M.~M.,
  {Weiss}, J.~W., {Porco}, C.~C., {Burns}, J.~A., 2010{\natexlab{b}}. {An
  analytic parameterization of self-gravity wakes in Saturn's rings}. \aj  139,
  492--503.

\bibitem[{{Ward}(1986)}]{Ward86}
{Ward}, W.~R., 1986. {Density waves in the solar nebula: Differential Lindblad
  torque}. \icarus  67, 164--180.

\bibitem[{{Ward}(1997)}]{Ward97}
{Ward}, W.~R., 1997. {Survival of planetary systems}. \apjl  482, L211--L214.

\end{thebibliography}
\end{document}